\documentclass[conference]{IEEEtran}
\ifCLASSINFOpdf
\else
\fi
\usepackage{graphicx}
\usepackage{subfigure}

\begin{document}
\title{Arthur: a new ECA that uses Memory to improve Communication}

\author{\IEEEauthorblockN{Paulo Knob, Willian S. Dias, Natanael Kuniechick, Joao Moraes, Soraia Raupp Musse}
\IEEEauthorblockA{Graduate Programa in Computer Science, School of Technology, Pontifical Catholic University of Rio Grande do Sul\\
Porto Alegre, Brazil\\
Email: soraia.musse@pucrs.br
}
}


%Telephone: (800) 555--1212, Fax: (888) 555--1212}


\maketitle

\begin{abstract}
%\boldmath
This article proposes an embodied conversational agent named Arthur. In addition to being able to talk to a person (using text and voice), he is also able to recognize the person he is talking to and detect his/her expressed emotion through facial expressions. Arthur uses these skills to improve communication with the user, also using his artificial memory, which stores and retrieves data about events and facts, based on a human memory model. We conducted some experiments to collect quantitative and qualitative information, which show that our model provides a consistent impact on users.
\end{abstract}


% \fi
\IEEEpeerreviewmaketitle

\section{Introduction}
\label{sec:introduction}

Human beings are the only known species that use spoken language to communicate, having a skill developed in communication that uses factors other than speech, such as, for example, body expressions and gazing~\cite{cassell2000embodied}.
%A concept that is studied as relevant during communication is empathy, which is the sharing of emotions between individuals, as well as the behavior of adopting another person's point of view~\cite{de2017mammalian}. For example, if someone is talking with a person who just lost a beloved relative, he/she can perceive this person is truly sad and also feel sadness as well.  %Other examples could be facial expression, memory, gesturing, and so on. 
In this context, Embodied Conversational Agents (ECAs) are virtual agents which are able to interact and talk with humans in a natural way. In the last years, many research have been conducted to improve the quality of the communication abilities of such ECAs, both verbal and non-verbal~\cite{yalccin2020empathy,biancardi2019computational,sajjadi2019personality}. A fair amount of effort has been focused on ECAs that help people to have a healthier life~\cite{kramer2019developing,spitale2020multicriteria,das2019generation}, in clinical interviews~\cite{philip2020trust,martinez2019assessment}, training specific skills~\cite{chetty2019embodied,ayedoun2019adding}, among others.

%The present work aims to propose an Embodied Conversational Agent (ECA) with general purpose, called Arthur, endowed with many abilities. 
The present work aims to propose an Embodied Conversational Agent (ECA) with general purpose, called Arthur, endowed with a memory module and many other abilities.
Besides a conversational module, using text and voice, this ECA is able to recognize the person he is talking to, as well to assess the user emotional state through his/her facial expressions. Also, Arthur is able to demonstrate different levels of emotion through his facial expressions. Lastly, Arthur is equipped with a memory module, which tries to replicate the behavior of human memory and, thus, allows for Arthur to learn information with and from the user while interacting. Moreover, it allows Arthur to remember such information later in the conversation or, even, in a different interaction.
%The empathy is build in the communication with the user in mainly three parts of our model: firstly, through a pre-defined module of communication where Arthur asks questions about the user (demonstrating interest in the conversation); then, in the module of memory once the user feels that Arthur remember him/her; and finally with simple facial expressions that Arthur applies as a result of detecting facial expression module of Arthur.

%JJ: Apresentação das outras seções
%PA: done
Section~\ref{sec:relatedWork} presents several work related with the model proposed in this work. Section~\ref{sec:method} details our methodology, while Section~\ref{sec:results} shows some results achieved when users know Arthur. Finally, Section~\ref{sec:conclusion} presents our final considerations.

\section{Related Work}
\label{sec:relatedWork}

This Section presents some work related with methodologies used to build Arthur. 
%Because the lack of space we referred more representative work for this paper. 
Section~\ref{sec:chatbots} presents two interesting researches developed in the context of
chatbots, while Section~\ref{sec:human-memory} presents papers that discuss the behaviour of human memory and its use for virtual agents. Finally, Section~\ref{sec:will-related} discusses the importance of appearance and expressiveness for virtual agents, also presenting work related with Embodied Conversational Agents (ECAs).

\subsection{Chatbots}
\label{sec:chatbots}

Zhou et al.~\cite{zhou2018emotional} focus their work on the proposal of the Emotional Chat Machine (ECM), a chatbot that can generate relevant responses to content that are also emotionally consistent. As discussed by the authors, there are three main challenges: obtaining high-quality data marked by emotions in a large-scale corpus, considering emotions in a natural and coherent way and incorporating that information of emotions into a neural model. The results achieved by their work show that ECM can generate appropriate responses, if the emotion category of the response and the emotion of the post both belong to one of the frequent Emotion Interaction Patterns (EIP). EIP is defined as the pair: categories of emotion of the publication and its response. 

Zhang et al.~\cite{zhang2019consistent} propose to solve the problem of consistency on chatbot responses, concerning both context and personas (casual speaker). In their work, they present a self-supervised approach that uses the natural structure of conversational data to learn and leverage both topic and persona features. For the tests, two data sets are used: Twitter FireHose, collected from 2012 until 2016; and Maluuba dataset. The results achieved indicate that the proposed model is able to capture meaningful topics and personas' features. Also, the incorporation of the learned features helps to significantly improve the quality of generated responses on both data sets, even when comparing with models which explicit persona information.

In our method, we mix a pre-defined dialog with an existent technology to provide chat and voice, in order to build the communication module of Arthur. More details are given in Section~\ref{sec:chat-voice}.

\subsection{Human and Agent Memory}
\label{sec:human-memory}

One of the most accepted models concerning human memory cited in literature is known as Autobiographical Memory. As defined by Bluck et al.~\cite{bluck1998reminiscence}, autobiographical memory is "a system that encodes, stores and guides retrieval of all episodic information related to our person experiences". Also, according to Conway et al.~\cite{conway2000construction}, autobiographical memory can be grouped in three levels: lifetime periods, general events and event-specific knowledge. So, such memories can be directly accessed if the cues are specific and relevant to the person. Otherwise, if the cues are too general, a generative retrieval process must be used to produce more specific cues for the retrieval of relevant memories. 

Following this definition of autobiographical memory, Wang et al.~\cite{wang2016modeling} build a model to mimic such behavior. Their model, known as Autobiographical Memory-Adaptive Resonance Theory (AM-ART), is a three-layer neural network that encode lifetime periods, general events and event-specific knowledge, respectively, being, therefore, consistent with the model presented by Conway et al.~\cite{conway2000construction}. Also, it encodes the 5W1H schema, which represents \textit{when} an event occurred, \textit{where} it happened, \textit{who} was involved, \textit{what} happened, \textit{which} pictorial memory was associated with the event and \textit{how} was the person feeling during the event. The results achieved by their work show that AM-ART was able to perform better that the keyword-based query method, since the last can not deal with noisy cues in many existing photo or memory repositories.

The main difference between the work presented in this Section and ours is that we do not train a neural network. We modeled our memory in a procedural way, similarly to~\cite{kope2013modeling}. Also, our memory retrieval can be guided by the emotional state and, even, change the mood of the agent. More details are presented in Section~\ref{sec:agent-memory}.

\subsection{Appearance and Expressiveness}
\label{sec:will-related}

%In Demeure et al~\cite{demeure2011believability}, the authors provided the evaluation of perception about virtual agents, which one of the main factors analysed was believably, which can be related either to physical features, emotions and personality of a virtual agent. The authors focused on how the use of the agent's appropriate emotional response, in relation to its context, could affect the perception of it. They also researched whether the use of verbal and non-verbal expressions has more impact than the use of just one or none. The results showed that an adequate emotional response can lead to a better perception of the characteristics of the agents mentioned.

When developing an interactive agent, there is a concern about how it presents itself. It is important that the interaction occurs so that the subject in contact with this agent does not feel uncomfortable and/or embarrassed. As shown in Dill et al.~\cite{dill2012evaluation}, when trying to present a character who pretends to be human, there is a certain strange feeling (Uncanny Valley) when it seems human, but not as close as expected. It should be noted that, considering the face of a character, the elements that can cause more strangeness are the eyes and the mouth.

Real human eyes tend to not stay static for too much time, having some involuntary or voluntary movement at some point. Such phenomena is known as saccade behavior, being defined by Leigh and Zee~\cite{leigh2015neurology} as rapid movements of both eyes from one gaze position to another. Lee et al.~\cite{lee2002eyes} propose an algorithm 
to simulate such behavior. The authors obtained a statistical model through the analysis of eye-tracking images from a subject during a conversation with a software. In combination with literature data, they achieved a model capable of generate movements considering various aspects, like magnitude and duration, of a human saccade. As result, the model obtained a natural and friendly perception from subject into author's survey.

In this work, we chose to model our virtual agent in a cartoon manner to avoid the strange feeling discussed by Dill et al.~\cite{dill2012evaluation}. Also, we included saccade eyes movement, following the algorithm proposed by Lee et al.~\cite{lee2002eyes}. More details are presented in Section~\ref{sec:facial-express}. 

\section{Proposed Model}
\label{sec:method}

This work aims to build an interactive empathetic ECA which can store and recover memories about previous interactions: Arthur. One of his features is being able to recognize the person he is talking to. Also, Arthur has a memory to store knowledge about people he interacts, such as their face, emotion and interactions. Finally, in order to keep the interactions as natural as possible, the agent is able to properly react to interactions with people, delivering meaningful responses and facial cues.
%, so providing empathy.

\subsection{Overview}
\label{sec:overview}

\begin{figure}[!htb]
  \centering
  \includegraphics[width=\linewidth]{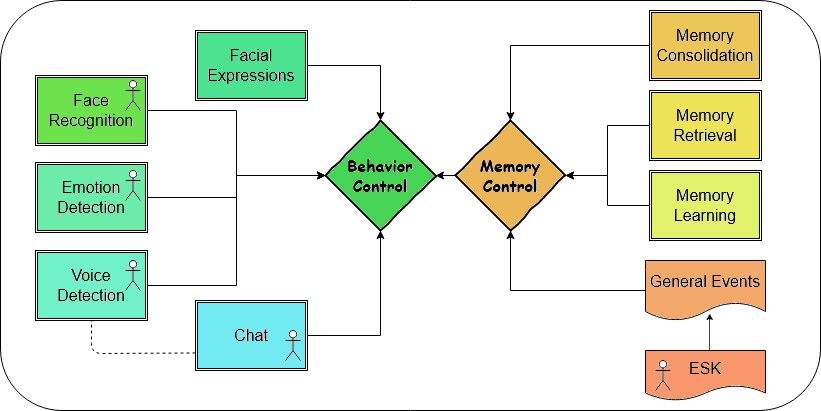}
  \caption{Overview of the proposed model. Behavior Control is responsible to define the appropriate behavior of the virtual agent, while the Memory Control is responsible of store and retrieve memories. The dolls inside the boxes represent the modules where an user input exist.}
  \label{fig:overview-model}
\end{figure}

The overview of Arthur is illustrated in Figure~\ref{fig:overview-model}. This work was mostly developed using Unity3D~\cite{Unity}. Some modules were developed using Python, as, for example, the Face Recognition module.

Behavior Control is responsible for defining Arthur's appropriate behavior. This module uses all the information available to Arthur, that is, the person recognized by Arthur and with whom he is interacting, the user's facial expressions, facts and information stored in Arthur's memory, the user's emotion detected and the dialogue through text/voice.

The Chat and Voice Detection modules are responsible by the interaction between the user and the agent, in the scope of verbal and textual behavior, where the Chat Module allows for the user to type some sentence and the Voice Detector Module is able to transform the voice of the user into words, which are used as input to the agent. The Face Recognition Module allows Arthur to recognize the person he is talking to, while the Emotion Detection Module allows him to identify the person's perceived emotion. The Facial Expressions Module is responsible to animate the facial expressions of Arthur, such as emotions and eyes movement. 
 
Finally, Memory Control is responsible for managing the memory of the virtual agent and is linked to all memory resources (i.e., memory learning, memory retrieval, memory consolidation, general events and ESK). These aspects will be explained in  Section~\ref{sec:agent-memory}.
We use a common method to define artificial memories that is to use Short and Long term memories, where information is transferred from one to the other (STM and LTM, respectively)~\cite{loftus2019human}. 
So, during the interaction between Arthur and the user, the information is stored in STM. Then, a phase called Memory Consolidation Module deals with the consolidation of the Long-Term Memory (LTM) of the agent. Therefore, saved memories can be forgotten by Arthur if their importance is too low or if they are not used for too long. Such process is detailed in Section~\ref{sec:memoryConsolidation}.

Next sections give more details about each of the modules.

\subsection{Chat and Voice Detector}
\label{sec:chat-voice}

The Chat module allows for the user to communicate with the virtual agent, by speaking or writing words and sentences.  The user can use its own voice to interact with the virtual agent or write sentences. The Voice Detection module is able to transform the voice input into words, which can be understood by the agent. To do so, the DictationRecognizer~\footnote{https://docs.unity3d.com/ScriptReference/Windows.Speech. DictationRecognizer.html} class was used. It is available for C\# and, thus, for Unity3D, allowing to access the system scripts which deal with voice translation to text. Such tool was used for the ease of use, once it is already integrated with Windows itself.

\subsubsection{Arthur Conversation}
\label{sec:conversation}

At the beginning of the conversation, we modeled some initial questions that our agent can ask the person it is talking to. These questions serve both as an ice-breaker and to know some basic information about that person, that are going to be stored in the memory. To the extent of this work, the modeled questions are:

\begin{itemize}
    \item How old are you?
    \item Do you work?
    \item Do you study?
    \item Do you have children?
\end{itemize}

Each question can, also, have another question(s) linked with them. For example, if the person answers that it has children, the virtual agent can ask how many children, or the names of them. All this information is stored in the memory of the agent, as further explained in Section~\ref{sec:agent-memory}. 
If, at any time, the user asks something about a certain subject that Arthur already knows, he can answer. For example, suppose that John tells Arthur that he is 42 years old. Then, in another conversation, someone asks the virtual agent "how old is John?". In this case, Arthur is able to answer that John is 42 years old. Finally, if Arthur is unsure about how to handle a question, the user's written/spoken sentence is sent to a chatbot API~\footnote{https://rapidapi.com} that processes the question and returns an answer, which is presented to the user. Since we are using an API for chatbot responses, it is quite easy to use another chatbot model, if necessary.

\subsection{Face Recognition}
\label{sec:face-recog}

The face recognition module allows Arthur to recognize the person he is talking to. To do this, we base our work on the Python library for face recognition~\footnote{Available at https://github.com/ageitgey/face\_recognition}. In summary, this library is a trained neural network that allows to recognize people's faces by comparing a particular image with other images saved in a data directory. If the person photographed has a low chance of being in the database (returned by the method), it is a new person, so it is stored in the dataset for further consultation. For more information, see the link in the footnote.

\subsection{Emotion Detection}
\label{sec:emotion-detect}

The Emotion Detection module provides to Arthur the skill to detect the emotion of the user. %person which is talking with the virtual agent and appearing at the webcam. In order to do so, w
We use the Affectiva~\footnote{https://developer.affectiva.com/} plugin, available for Unity. Affectiva is a well accepted framework which uses a trained neural network to identify many points on the person's face. Depending on the arrangement of such points, the network is able to detect the emotion that the person is expressing. For more information, see the link in the footnote.

\subsection{Agent Memory}
\label{sec:agent-memory}

Arthur's memory is used to record data collected during interactions with people. For this, we adopted a model known as Autobiographical Memory. In short, as defined by Bluck et al.~\cite{bluck1998reminiscence}, autobiographical memory is "a system that encodes, stores and guides retrieval of all episodic information related to our person experiences". Therefore, it is capable of storing different types of information (for example, text, episodes, images, voice) and retrieving it, if a certain suggestion (or suggestions) is informed. 
For example, when someone meets a new person, they usually greet each other and say their names. If the memory serves that person well, he/she will be able to remember the other person's name when they meet again.

According to Conway et al.~\cite{conway2000construction}, autobiographical memory can be grouped in three levels: Lifetime Periods, General Events and Event-Specific Knowledge. Lifetime Periods serve as an index to cue General Events and can be linked to different periods across the lifespan of a person. For example, someone's first job can be seen as a lifetime period. General Events are events that occur within a certain Lifetime Period. Using the first example of work given earlier, some general events within this can be the first day of work, meeting a new colleague or a happy hour with people in the office. Finally, the Event Specific Knowledge (ESK) includes a set of resources that form the memory of a given event. This information can be stored in different types, such as images, grammatical or audio.

As it is possible to see in Figure~\ref{fig:overview-model}, we model our memory using General Events and ESK. As the intent of our virtual agent is to interact with people, we chose not to model Lifetime Periods. General Events represent the events which occur during the interaction between Arthur and the user. For example, when Arthur meets someone new, a new General Event is generated (i.e. Meet new person). More details about General Events are given in Section~\ref{sec:generalEvents}. Moreover, following the autobiographical memory model, General Events are comprised of a set of resources known as ESK. In this work, we store such resources in two levels: Short Term Memory (STM) and Long Term Memory (LTM). More details are given in Section~\ref{sec:stmAndLtm}.

\subsubsection{General Events}
\label{sec:generalEvents}

In our model, General Events are comprised of:

\begin{itemize}
    \item Timestamp: the moment this event was created or updated.
    \item ID: an unique ID which refers to this event.
    \item Type: refers to the type of the event and there are currently three possible types: \textit{i)} Meet new person: when the agent meets someone new; \textit{ii)} Learn thing: when the agent learn something new; and \textit{iii)}  Interaction: when it does not fit in any of the other types).
    \item Emotion: the emotion associated with the event, recognized from the Emotion
    Detection module. For example, if a person is talking about the death of its beloved pet, it is expected that such person assumes a sad face. So, this general event should be associated with the Sadness emotion.
    \item Polarity [-1,1]: the polarity of the sentence, it means, if it is a positive or negative sentence. For example, the sentence "I am feeling good" would be a positive sentence, while the sentence "I am not feeling good" would be a negative one. When we divide the sentence into tokens, we also use the sentiment library\footnote{https://www.nltk.org/howto/sentiment.html}, provided by NLTK, to calculate the polarity of the given sentence. To do so, we chose to work with Vader method. It returns a float value lying between -1 and 1, where negative values reflect a negative polarity and positive values reflect a positive polarity.
    \item Resources: a list with the resources (i.e. ESK) associated with this General Event. Using the previous example, some resources associated with this event could be a picture of the deceased pet and some grammatical information like the pet name, death and when it passed away.
\end{itemize}

\subsubsection{ESK: STM and LTM}
\label{sec:stmAndLtm}

As commented before, we store the resources (i.e. ESK) in two levels: STM and LTM. According to Loftus et al.~\cite{loftus2019human}, STM is used to store important information for a short period of time, while LTM is an information storage with virtually unlimited capacity that each human being has. In our work, we model STM and LTM separately. The STM is comprised of a list of resources. This list of resources can have, at most, seven items, as defined by Miller's Law~\cite{miller1956magical}. If a new resource should enter the STM, the less important information is forgotten according its weight (i.e. the resource with the lower weight is removed). Each resource  has the following information:

\begin{itemize}
    \item Timestamp: the moment this resource was created or updated.
    \item ID: an unique ID which refers to this resource.
    \item Type: refers to the type of the resource (e.g. grammatical, image, audio).
    \item Information: refers to the information itself that the resource should reflect. For example, if the resource is an image, the information is going to contain the path to such image.
    \item Activation [0,1]: As described in Loftus et al.~\cite{loftus2019human}, an information present in the short-term memory is rapidly forgotten (around 15 seconds), unless it is rehearsed, it means, repeated over and over. The activation represents this rehearsal process. When a new resource enters the STM of Arthur, activation is set at its maximum value (i.e. 1). As time passes by, a logarithmic-based decay function is applied for each resource, diminishing their activation value. It is defined as follows: $A_{ID_{STM}}^* = Log(A_{ID_{STM}} + 1)$, where $A_{ID_{STM}}^*$ is the new activation value of STM for resource $ID$, $A_{ID_{STM}}$ is the current activation value and $Log()$ is a logarithmic function. If any resource in the memory is rehearsed (e.g. remembered, seen again), its activation value is set back to 1. This attribute exists only for STM.
    \item Weight [0,1]: represents the importance of the resource. For example, meeting a new person can be considered more important than talking about the weather. We empirically defined the initial importance of each resource (at STM) based on the Type of the General Event it belongs to, as follows: Meet new person: 0.9; Learn thing: 0.9; Interaction: 0.1. At LTM this attribute is not initialized but indeed affected by the values of $W_{ID_{LTM}}^* = f(A_{ID_{STM}}^*,W_{ID_{STM}}^*)$ at STM, in the consolidation of the memory, as explained in Section~\ref{sec:memoryConsolidation}.
\end{itemize}

In order to store grammatical resources, we divide a given sentence in significant tokens and keep each of them separately. To do so, we use the Natural Language ToolKit (NLTK)\footnote{https://www.nltk.org/}. since it is a well accepted tool for natural language processing. It is a platform for building Python programs to work with human language data. Such tool is able to "tokenize" a text, it means, split it in sentence or word tokens. It is also able to remove "stop words" from the text, it means, words which have low or no meaningfulness for the context of the sentence. We built a script which is able to remove the "stop words" from the text and "tokenize" it word by word. For example, assuming the user told to the virtual agent "I am going on vacation with my dad to Glasgow", the tokens returned by the NLTK script would be "vacation", "dad" and "Glasgow".

The same type of resources can be stored in STM and LTM, but the LTM is theoretically unlimited, it means, can maintain an unlimited number of resources.
To keep track of the content of the Long-term memory, we save all its information in a database, so we can retrieve it at any moment.

\subsubsection{Memory Retrieval}
\label{sec:memoryRetrieval}

Assuming the information is stored, how can one retrieve such information if required? The Memory Retrieval module deals with the retrieval of the information from the storage. According to Conway et al.~\cite{conway2000construction}, there are two types of memory construction: Generative Retrieval and Direct Retrieval. Generative retrieval is a method guided by cues, it means, the retrieval depends on some "hint" in order to find some information. For example, the word "dog" can make some people remember of its beloved pet. Thus, the word "dog" acted like a cue to the generative retrieval method, constructing a memory of a special dog. In its turn, Direct Retrieval method refers to  spontaneously recalled memories, it means, memories that are recalled automatically, with no apparent cue. According to Berntsen~\cite{berntsen1996involuntary}, such process occurs between two to three times each day. In our work, the Memory Retrieval is developed following the Generative Retrieval method. When the user interacts with the virtual agent, the information provided can be used as cue(s) to the Retrieval method. For example, if the user tells "I went fishing with my dad", the words "fish" and "dad" are used as cues for the retrieval method, so we proceed with a search of those words in the database.

\subsubsection{Memory Consolidation: from STM to LTM}
\label{sec:memoryConsolidation}

This module is responsible for consolidating the Long-term Memory of the virtual agent based on data available at STM. 
According Klinzing et al.~\cite{klinzing2019mechanisms}, the formation of such LTM is a major function of sleep. As their work affirms, the authors "consider the formation of long-term memory during sleep as an active systems consolidation process that is embedded in a process of global synaptic downscaling". In short, the consolidation process prioritizes important memories over mundane ones. For example, emotional memories are more important and have more impact than neutral memories. Therefore, less important information can be placed in a second plan or, even, forgotten.

In our work, we developed a module to simulate such process, which uses two information from the STM's resources: Activation and Weight. If the Activation value $A_{ID}$ of a given resource $ID$ is below an empirically defined threshold (i.e. $A_{ID}<0.2$), the Weight attribute of this resource is reduced at LTM, as follows: $W_{ID_{LTM}}^* = Log(W_{ID_{STM}} + 1)$, where $W_{ID_{LTM}}^*$ is the new Weight this resource will have at LTM, $W_{ID_{STM}}$ is the current weight at STM, and $Log()$ is a logarithmic function. Such process is repeated for all stored resources at STM. Then, resources which have low importance are wiped out from the LTM. We empirically define that a resource has low importance if its weight drops below 0.2. In its turn, General Events are not forgotten by the virtual agent, unless all the resources belonging to a given event are also forgotten. We do so to make possible for the agent to forget just parts of the information, just like it happens with a real person. In addition, please notice that during the memory consolidation, all data at STM is erased.

\subsection{Facial Expressions}
\label{sec:facial-express}

In order to model our agent facial characteristics, we proceeded the following way: the various parts of agent's face (such as eyes, mouth, etc) where created using Unity3D sprites. Then, after placing each of them into their position, a set of constraints were defined in order to avoid further distortions, e.g., when agent's head moves, the other parts follow that movement in a proper way. Finally, the different expression where modeled using a plug-in called Anima2D~\footnote{\label{foots}https://assetstore.unity.com/packages/essentials/unity-anima2d-79840}.

The processes of animation consists in the usage of skeletal animation. For a given mesh in the face that should suffer some kind of deformation, such as the eyebrows and mouth, a set of bones is defined and linked to this mesh. After that, it is possible for us to animate the agent through the time just by applying transformations into the bones, which is done through Anima2D\footnote{See footnote 7}. Therefore, after animating a certain facial expression (which can by either animated or a static one), we can save it for further usage. The expressions of our agent are: anger, disgust, doubt, fear, joy, sadness, surprise and worry. Also, there is a neutral expression and one for sleeping (in order to indicate that the memory consolidation process was triggered).

Finally, in order to give some expression to the eyes of the virtual agent, it was implemented saccade movements on them, according to Lee et al~\cite{lee2002eyes}. As commented in Section~\ref{sec:will-related}, human eyes are not static for much time, tending to have involuntary movements, even when the person is focusing its vision on something. Such phenomena is known as saccade, from the french word \textit{saccade}, which means \textit{jerk}. As defined by Leigh and Zee~\cite{leigh2015neurology}, saccades are rapid movements of both eyes from one gaze position to another. 

% \subsection{Arthur's Empathy}

% Empathy can involve cognitive attributes or affective attributes which also can be combined~\cite{goldstein1985empathy}. Cognitive attributes of empathy involve cognitive reasoning used to understand another person's experience~\cite{hojat2007empathy}. Emotional or affective attributes involve physiological enthusiasm and spontaneous affective responses to someone else's display of emotions~\cite{prendinger2005using}. With these definitions, we propose a simple empathetic behavior for Arthur, using a metaphor of mimicry reaction. i.e., the ability to detect and recognize emotions being able to mime the user's emotion (interaction metaphor of mimic). So, Arthur can mimicry the emotions and also look directly to user he is interacting. For this, we calculated the mean point coordinates from the input image and remap them to the coordinates of Unity by identifying the minimum and maximum values of the both universes and converting the coordinates based on these bounds. When this option is selected, a text component is shown on the interface with the current emotion detection. Additionally, Arthur mimics the user expressions, activating the animations corresponding to the neural network's output, emphasizing his empathetic properties.

\section{Results}
\label{sec:results}

With all features presented in Section~\ref{sec:method} working altogether, we developed an Embodied Conversational Agent (ECA), named Arthur, able to recognize the person he is talking to, as well the emotions hinted by this person's facial expressions.Arthur is also capable of demonstrate some level of emotions, depending the situation. Also, it can learn things from the user and remember such information when needed.

\begin{figure}[!htb]
  \centering
  \includegraphics[width=\linewidth]{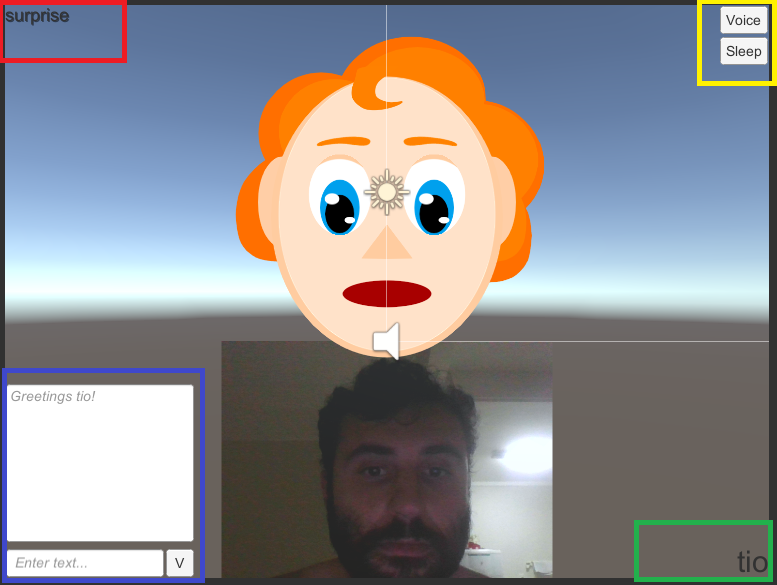}
  \caption{Arthur and its interface. The green square is where the name of the recognized person is presented. The blue square is where the person can chat with Arthur. The red square is where the person's emotion is shown. Finally, in the yellow square we have the Voice and Sleep buttons, which allow for the user to talk naturally with Arthur and puts it to sleep, respectively.}
  \label{fig:arthur}
\end{figure}

Figure~\ref{fig:arthur} shows Arthur and its interface. We chose to model Arthur in a cartoon manner to avoid the Uncanny Valley, as commented in Section~\ref{sec:will-related}. At the bottom, the image captured by the camera is shown, with the face of the person talking with Arthur. The green square is where the name of the recognized person is presented. The blue square is where the person can chat with Arthur. The larger text area is where Arthur speaks, while the smaller input text below is where the user can write his/her sentences. The red square is where the person's emotion is shown. Finally, in the yellow square, two buttons: Voice and Sleep. The Voice button toggles the voice management, allowing for the user to talk with Arthur using its own voice, while the Sleep button puts Arthur to sleep, triggering the memory consolidation process, as explained in Section~\ref{sec:memoryConsolidation}.

In order to test our method, we developed some experiments exploring the various features of Arthur. Section~\ref{sec:result-agent-memory} shows some emergent behaviors regarding Arthur's memory while Section~\ref{sec:results-user-study} presents research carried out with subjects on the functioning and behavior of Arthur. 

\subsection{Agent Memory}
\label{sec:result-agent-memory}

In order to evaluate the Arthur's memory, we test some possible scenarios:

\begin{itemize}
    \item Introduction Scenario: Arthur meets a new person and starts to asking questions about him/her. After that, we check if Arthur can remember all information.
    \item Learning Scenario: we feed Arthur with information about some objects. Later, we check if it recognizes such objects.
\end{itemize}

\subsubsection{Introduction Scenario}
\label{sec:result-intro-scenario}

\begin{figure}[!htb]
    \centering
    \subfigure[Arthur does not know this user.]{\includegraphics[width=0.49\linewidth]{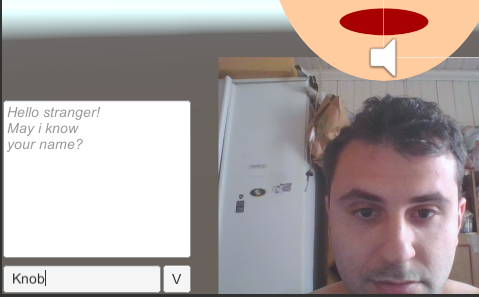}}\hfill
    \subfigure[After introduction, Arthur is able to remember this person.]{\includegraphics[width=0.49\linewidth]{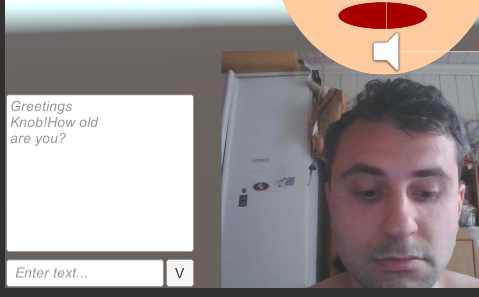}}
    \caption{Arthur is meeting a new person. In (a), he asks for this user's name. 
    After a proper introduction, Arthur stores that person's name in his memory and then he can remember that person's face and name, greeting him/her when they see each other again (b).}
    \label{fig:meet-new-person}
\end{figure}

In this scenario, Arthur begins with no knowledge about the person it is seeing on the webcam, thus, its first action should be to ask who it is. Figure~\ref{fig:meet-new-person} shows how Arthur deals with this situation. In Figure~\ref{fig:meet-new-person} (a), Arthur is meeting a new person and, therefore, asks for this person's name with the statement "Hello stranger! May I know your name?". After the proper introduction (i.e. the person in question answers that its name is "Knob"), Arthur stores the name of this person on its memory. When they meet again (Figure~\ref{fig:meet-new-person} (b)), Arthur is able to remember both the face and the name of this person. Then, it proceeds to a cordial greeting ("Greetings Knob!"). Plus, Arthur tries to maintain the conversation asking questions about the person. In Figure~\ref{fig:meet-new-person} (b), it is possible to see Arthur asking the age of the person ("How old are you?"). When the answer is given, such information is also stored into the memory, increasing the information Arthur knows about this user.

\subsubsection{Learning Scenario}
\label{sec:result-learning-scenario}

\begin{figure}[!htb]
    \centering
    \subfigure[Arthur asking a question to the person.]{\includegraphics[width=0.49\linewidth]{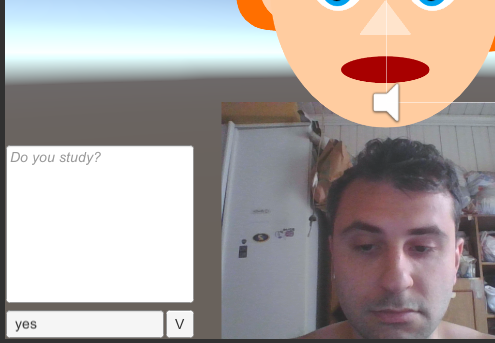}}
    \subfigure[After being asked about the age of Knob, Arthur gives the answer.]{\includegraphics[width=0.49\linewidth]{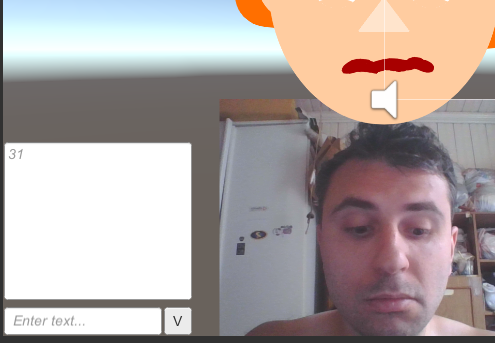}}
    \subfigure[After being asked about if it knows a cellphone, Arthur says it does not.]{\includegraphics[width=0.49\linewidth]{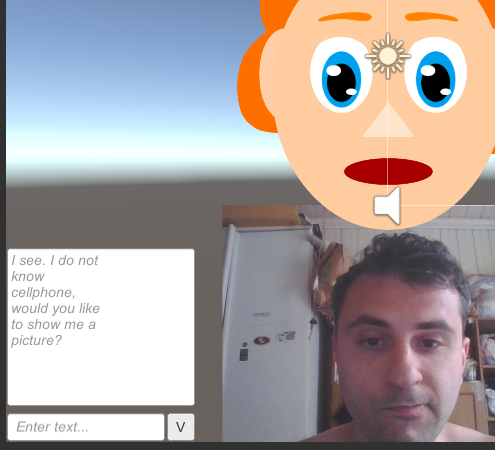}}
    \subfigure[After it learns what a cellphone is, Arthur is able to identify it.]{\includegraphics[width=0.49\linewidth]{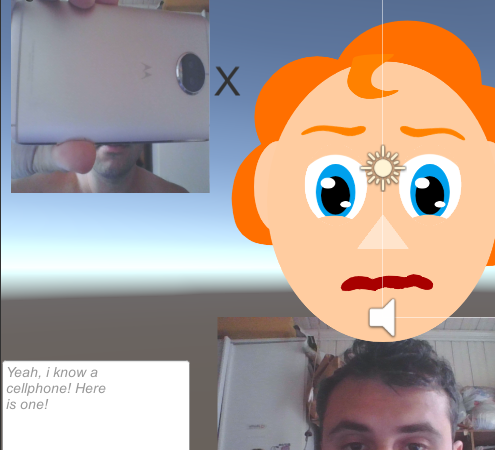}}
    \caption{Arthur learning new things. In (a) and (b), it is shown the process where Arthur learns new information about a given person. In (c) and (d), Arthur learns what a cellphone is.}
    \label{fig:learning-things}
\end{figure}

As commented in Section~\ref{sec:agent-memory}, Arthur is able to learn information from the person he is talking to. Figure~\ref{fig:learning-things} shows Arthur learning 2 types of information. As commented in Section~\ref{sec:result-intro-scenario}, Arthur tries to starts or maintain a conversation with the person he is interacting, e.g., asking questions about him/her. Figure~\ref{fig:learning-things} (a) shows Arthur asking if the person studies. All possible questions are presented in Section~\ref{sec:chat-voice}. In Figure~\ref{fig:learning-things} (b), Arthur answers the question about the age of Knob, saying he is 31 years old. It is able to do so because the person (i.e. Knob) already informed Arthur about his age in a past interaction.

Besides learning information about the person Arthur is talking to, he can also learn about new things. For example, in Figure~\ref{fig:learning-things} (c), it is asked if Arthur knows what a cellphone is. "No, it does not!", is the Arthur answer. Then, he asks the person if he/she desires to show him a picture of a cellphone. If the person agrees, a new term is stored in Arthur's memory, next to the image provided. Then, when Arthur is asked again about a cellphone (Figure~\ref{fig:learning-things} (d)), he is able to remember what it is, showing the image which was provided before.

\subsection{Study with Subjects}
\label{sec:results-user-study}

In order to test our hypothesis with a qualitative evaluation, we conducted an experiment with subjects. Each person was presented to three video sequences. In the first video sequence, Arthur meets a new person and asks questions about he/she, as explained in Section~\ref{sec:chat-voice}, but the Memory Module is deactivated, thus, it does not retain information. In the second video sequence, the same interaction is conducted, but the Memory Module is activated, thus, Arthur can retain information and remember it further. Finally, in the third video sequence is specifically presented the process where Arthur learns a new information and remembers it further. Arthur is asked if he knows an object (i.e. cellphone), which he does not. Then, we show to him what a cellphone is and ask him again if he knows a cellphone. Since the Memory Module is activated, Arthur remembers what a cellphone is and answer the person. After each video sequence, each subject was asked to answer some questions:

\begin{itemize}
    \item Q1: How do you evaluate your comfort level regarding agent's appearance?
    \item Q2: How do you evaluate the agent's facial expressions realism level?
    \item Q3: How do you evaluate agent's information comprehension abilities?
    \item Q4: After watching the interaction, how do you evaluate agent's memory?
\end{itemize}

Subjects answered each question following a Likert scale, where: 

%\begin{itemize}
    %\item 
    $1$: Very low,
   % \item 
    $2$: Low,
    %\item 
    $3$: Moderate,
    %\item 
    $4$: Good and
   % \item 
    $5$: Very good.
%\end{itemize}

Regarding our expectations, we formulate the following hypothesis:
%SO: aqui tem que dizer se 1 era fracamente de acordo, mas tem que especificar o que era cada nível
%PA2: adicionado

%PA: se tiver mais hipoteses ou precisar alterar alguma...
\begin{itemize}
    \item H1: Since the facial expressions 
    %and the empathetic behavior 
    are the same for all three videos, we expect little or no difference of how people will evaluate both their comfort level and the realism level of the agent's facial expressions (Q1 and Q2), throughout the video sequences.
    \item H2: We believe that the memory of Arthur is going to help him to understand the context of the interaction better. Therefore, in the videos where the Memory Module is activated, we believe the participants are going to rate Arthur's comprehension abilities better (Q3).
    \item H3: We expect that people rate Arthur's memory higher in Videos 2 and 3, when compared with Video 1 (Q4).
\end{itemize}

Concerning the subjects, we had 51 answers, where 39 were male and 12 were female. Also, we had participants ranging from 17 to 61 years old. The average age was 31.49 with a standard deviation of 12.49. All participants agreed with the terms of the research and conducted the experiment until the end.

\begin{table}[htb]
\centering
\caption{Quantity of answers acquired by the experiment, separated by each video sequence. The head numbers (i.e. 1, 2, 3, 4 and 5) represent each value of the Likert scale.}
\label{tbl:answers-experiment}
\begin{tabular}{|c||c||c||c||c||c|}
\hline
\textbf{Video 1} & \textbf{1} & \textbf{2} & \textbf{3} & \textbf{4} & \textbf{5} \\ \hline
\textbf{Q1} & 4 & 7 & \textbf{19} & \textbf{19} & 2 \\
\textbf{Q2} & 8 & 12 & \textbf{19} & 10 & 2 \\
\textbf{Q3} & 3 & 14 & \textbf{15} & \textbf{15} & 4 \\
\textbf{Q4} & 7 & 8 & \textbf{17} & 15 & 4 \\
\hline
\end{tabular}

\begin{tabular}{|c||c||c||c||c||c|}
\hline
\textbf{Video 2} & \textbf{1} & \textbf{2} & \textbf{3} & \textbf{4} & \textbf{5} \\ \hline
\textbf{Q1} & 6 & 9 & 15 & \textbf{19} & 2 \\
\textbf{Q2} & 5 & 12 & \textbf{20} & 12 & 2 \\
\textbf{Q3} & 4 & 4 & 14 & \textbf{23} & 6 \\
\textbf{Q4} & 5 & 4 & 11 & \textbf{22} & 9 \\
\hline
\end{tabular}

\begin{tabular}{|c||c||c||c||c||c|}
\hline
\textbf{Video 3} & \textbf{1} & \textbf{2} & \textbf{3} & \textbf{4} & \textbf{5} \\ \hline
\textbf{Q1} & 4 & 7 & 16 & \textbf{19} & 5 \\
\textbf{Q2} & 5 & 12 & \textbf{19} & 10 & 5 \\
\textbf{Q3} & 2 & 6 & 12 & \textbf{19} & 12 \\
\textbf{Q4} & 1 & 5 & 13 & \textbf{19} & 13 \\
\hline
\end{tabular}
\end{table}

Table~\ref{tbl:answers-experiment} shows the quantity of answers obtained per question, for each video sequence. In order to investigate the hypothesis variation of response of the users, we rely on ANOVA test. The results reveal an uniformity in the participants answers concerning Q1 (F(3.05)=0.63,p=.53)) and Q2 (F(3.05)=0.64,p=.52)), thus, there is little variation on the answers about the appearance and realism level of the agent (i.e. Q1 and Q2), which confirms H1. For Q1, most answers were between Moderate (3) and Good (4). For Q2, the most answers were Moderate (3). With this, we can also conclude that the appearance of Arthur has some space to improve, specially concerning 
%the empathetic behavior and 
its facial expressions.

An unexpected behavior was observed on the answers about the comprehension abilities and memory of the agent (i.e. Q3 and Q4). Both questions had a significant amount of answers between Moderate (3) and Good (4) for Video 1. For example, if we take Q4, it was expected to have most answers lying between Very Low (1) and Low (2), since the agent has the Memory Module deactivated. However, we had 17 answers for Moderate (3) and 15 for Good (4), while having a total of 15 for the ones we were expecting. One possible explanation could be the order of the sentences that Arthur speaks. Since they are presented to the user in an orderly way, it may cause the impression that Arthur is indeed understanding and processing the information given by the user. Further investigation would be necessary to confirm or disprove it.

Even with such unexpected behavior, it seems that H2 and H3 were also confirmed. In Video 1, we had a total of 19 answers for Good (4) and Very Good (5), for both Q3 and Q4. In Video 2, with the Memory Module activated, we had a total of 29 answers, in the same range, for Q3 and 31 answers for Q4. In Video 3, we had 31 for Q3 and 32 for Q4. Also, if we look at the amount of answers Very Low (1) and Low (2), we can clearly perceive that it diminishes in Videos 2 and 3, when compared with Video 1. Finally, the results of the ANOVA test show a change of perception when comparing Video 1 and Video 3, for both Q3 (F(3.93)=7.57,p=.007)) and Q4 (F(3.93)=11.31,p=.001)).

\section{Final Considerations}
\label{sec:conclusion}

This work presented a model of an Embodied Conversational Agent (ECA) endowed with a memory module and many other abilities, like face recognition, emotion detection and expressiveness. We conducted some experiments in order to test our model and collect both quantitative and qualitative information. The results achieved seem to confirm that Arthur presented the expected behavior.  Concerning the 
%empathy of the agent, as well its 
appearance of the virtual agent, despite having a good result in the user study, we conclude that it has room to improve. A few changes could be made regarding the geometric model and the animations, pursuing a better comfort level and higher realism. Also, concerning the memory model, a satisfactory result was achieved. In both quantitative and qualitative tests, the results achieved showed that the memory worked as intended and its influence on the interaction could be perceived. 

As for future work, there are a few paths that can be followed. As have already been commented, we plan to improve the appearance 
%and the empathy 
of the agent. Also, according the literature, as presented in Sections~\ref{sec:human-memory} and~\ref{sec:agent-memory}, the human memory is able to store information in different types, like text, images and voice. This work presents a memory model which deals with text and images, thus, the next logical step would be to make it able to deal with audio also. Moreover, besides being able to recognize the person it is talking with, the virtual agent could be able to recognize objects present in the image, both to learn what such object is and identify it later.

\section*{Acknowledgments}

The authors would like to thank to Brazilian agencies CNPq e CAPES for partially funding this project.

% \section*{Acknowledgment}




%\bibliographystyle{IEEEtran}


% \end{thebibliography}
\bibliographystyle{IEEEtran}
\bibliography{IEEEfull}

\end{document}